\documentclass[aoas,MSNbibl,nameyear,seceqn,dvips]{arximspdf}
\usepackage{dcolumn}

\doi{10.1214/12-AOAS583} %kopijuoti is PTS
\volume{7}
\issue{1}
\pubyear{2013}
\firstpage{295}
\lastpage{318}

\makeatletter
\newcolumntype{d}[1]{D{.}{.}{#1}}
\newtheorem{thmm}{Theorem}
\newtheorem{cor}{Corollary}[thmm]
\newproclaim{cond}{Condition}
\newproclaim{com}{Remarks}
\newcommand{\ov}{\overline}
\newcommand{\hatw}{\widehat}
\newcommand{\mb}{\mathbf}
\newcommand{\goto}{\rightarrow}
\newcommand{\var}{\operatorname{var}}
\newcommand{\cov}{\operatorname{cov}}
\makeatother

\begin{document}
\begin{frontmatter}

\title{Agnostic notes on regression adjustments to experimental data:
Reexamining Freedman's critique}
\runtitle{Regression adjustment}

\begin{aug}
\author{\fnms{Winston} \snm{Lin}\corref{}\ead[label=e1]{Linston@gmail.com}}
\runauthor{W. Lin}
\affiliation{University of California, Berkeley}
\address{Department of Statistics\\
University of California, Berkeley\\
Berkeley, California 94720-3860\\
USA\\
\printead{e1}} %adresu isvedimo komanda gale!
\end{aug}

% HISTORY:
\received{\smonth{10} \syear{2011}}
\revised{\smonth{7} \syear{2012}}

% ABSTRACT
%
\begin{abstract}
Freedman [\textit{Adv. in Appl. Math.} \textbf{40} (2008)
180--193;
\textit{Ann. Appl. Stat.} \textbf{2} (2008) 176--196]
critiqued ordinary least squares
regression adjustment of estimated treatment effects
in randomized experiments,
using Neyman's model for
randomization inference.
Contrary to conventional wisdom,
he argued that adjustment can lead to
worsened asymptotic precision,
invalid measures of precision, and
small-sample bias.
This paper shows that in sufficiently large samples,
those problems are either minor or easily fixed.
OLS adjustment cannot hurt asymptotic precision
when a full set of treatment--covariate interactions is included.
Asymptotically valid confidence intervals
can be constructed with the Huber--White sandwich standard
error estimator.
Checks on the asymptotic approximations are illustrated with
data from Angrist, Lang, and Oreopoulos's
[\textit{Am. Econ. J.: Appl. Econ.} \textbf{1:1} (2009) 136--163]
evaluation of strategies to improve college students' achievement.
The strongest reasons to support Freedman's preference for unadjusted
estimates are transparency and the dangers of specification search.
\end{abstract}

% KEYWORDS
%

\end{frontmatter}

%s1 #&#
\section{Introduction}\label{sec1}
One of the attractions of randomized experiments is that, ideally,
the strength of the design reduces the need for statistical modeling.
Simple comparisons of means can be used to estimate the average
effects of assigning subjects to treatment.
Nevertheless, many researchers use linear regression models to adjust for
random differences between the baseline characteristics of the
treatment groups.
The usual rationale is that adjustment tends to improve
precision if the sample is large enough and the covariates are
correlated with the outcome;
this argument, which assumes that the regression model is correct,
stems from \citet{Fisher32} and is taught to applied researchers in many fields.
At research firms that conduct randomized experiments
to evaluate social programs,
adjustment is standard practice.\footnote{\citet{Cochran57},
\citet{CoxM}, \citet{Raud}, and \citet{Klar}
discuss precision improvement.
\citet{Greenberg} document the use of regression adjustment
in many randomized social experiments.}

In an important and influential critique,
Freedman (\citeyear{Freed08a}, \citeyear{Freed08b}) analyzes the behavior of ordinary least squares
regression-adjusted
estimates without assuming a regression model.
He uses Neyman's (\citeyear{Ney23}) model for randomization inference:
treatment effects can vary across subjects, linearity is not assumed,
and random assignment is the source of variability in estimated
average treatment effects.
Freedman shows that
(i) adjustment can actually worsen asymptotic precision,
(ii) the conventional OLS standard error estimator is
inconsistent,
and (iii) the adjusted treatment effect estimator has a
small-sample bias.
He writes [\citet{Freed08a}],
``The reason for the breakdown is not hard to find: randomization does
not justify the assumptions
behind the OLS model.''

This paper
offers an alternative perspective.
Although I agree with Freedman's (\citeyear{Freed08b}) general advice
(``Regression estimates \ldots should be deferred until rates and
averages have been presented''),
I argue that in sufficiently large samples,
the statistical problems he raised
are either minor or easily fixed.
Under the Neyman model with Freedman's regularity conditions, I~show that
(i) OLS adjustment cannot hurt asymptotic precision when a full set of
treatment $\times$ covariate interactions is included,
and
(ii) the Huber--White sandwich standard error estimator is
consistent or asymptotically conservative (regardless of whether the
interactions
are included).
I~also briefly discuss the small-sample bias issue and the distinction
between unconditional and conditional unbiasedness.

Even the traditional OLS adjustment
has
benign large-sample properties when subjects are randomly assigned to
two groups of equal size. \citet{Freed08a} shows that in this case,
adjustment (without interactions) improves or does not hurt asymptotic
precision, and
the conventional standard error estimator is consistent or
asymptotically conservative. However, Freedman and many excellent
applied statisticians in the social sciences have summarized his
papers in terms that omit these results and emphasize the dangers of
adjustment.
For example, \citet{Berk} write:
``Random assignment does not justify any form of regression with
covariates.
If regression adjustments are introduced nevertheless, there is likely
to be bias in any estimates of treatment effects and badly biased
standard errors.''

One aim of this paper is to show that such a negative view is not
always warranted. A second aim is to help provide a more intuitive
understanding of the properties of OLS adjustment when the
regression model is incorrect.
An ``agnostic'' view of regression
[\citet{AngristImbens}, Angrist and Pischke (\citeyear{MHE}), Chapter~3]
is adopted here: without taking the regression model literally,
we can still make use of properties of OLS
that do not depend on the
model assumptions.

%s1.1 #&#
\subsection{Precedents}
Similar results on the asymptotic precision of OLS adjustment with interactions
are proved in interesting and useful papers by
\citet{Yang}, \citet{Tsiatis}, and \citet{Schochet},
under the assumption that
the subjects are a random sample from an infinite
superpopulation.\footnote{Although Tsiatis et al. write that
OLS adjustment \textit{without} interactions ``is
generally more precise than \ldots the difference in sample means''
(page~4661),
Yang and Tsiatis's asymptotic variance formula
correctly implies
that this adjustment may help or hurt precision.}
These results are not widely known, and Freedman was apparently
unaware of them.
He did not analyze
adjustment with interactions, but conjectured, ``Treatment by
covariate interactions can probably be accommodated too''
[Freedman (\citeyear{Freed08b}), page~186].

Like Freedman, I use the Neyman model, in which random
assignment of a finite population is the sole source of randomness;
for a thoughtful philosophical discussion of finite- vs.
infinite-population inference;
see Reichardt and Gollob [(\citeyear{Reichardt}), pp.~125--127].
My purpose is not to advocate finite-population inference, but to show
just how little needs to be changed to address Freedman's major
concerns.
The results may help researchers understand why and when
OLS adjustment can backfire.
In large samples,
the essential problem is omission of treatment $\times$ covariate
interactions,
not the linear model. With a balanced two-group design,
even that problem disappears asymptotically,
because two wrongs make a right
(underadjustment of one group mean cancels out
overadjustment of the other).

Neglected parallels between regression adjustment in experiments and
regression estimators in survey sampling turn out to be very helpful
for intuition.

%s2 #&#
\section{Basic framework}\label{sec2}
For simplicity, the main results in this paper assume a
completely randomized experiment with two treatment groups
(or a treatment group and a control group), as in \citet{Freed08a}.
Results for designs with more than two groups are
discussed informally.

%s2.1 #&#
\subsection{The Neyman model with covariates}
The notation is adapted from \citet{Freed08b}.
There are $n$ subjects, indexed by $i = 1, \ldots,
n$. We assign a simple random sample of fixed size $n_A$ to treatment $A$
and the remaining $n - n_A$ subjects to treatment $B$.
For each subject, we observe an outcome $Y_i$ and
a row vector of covariates $\mb{z}_i = (z_{i1}, \ldots, z_{iK})$,
where $1 \leq K < \min(n_A, n - n_A) - 1$.
Treatment does not affect the covariates.

Assume that each subject has two potential outcomes
[\citet{Ney23},
Rubin (\citeyear{Rubin74}, \citeyear{Rubin05}), \citet{Holland}], $a_i$ and $b_i$, which would be
observed under treatments $A$ and $B$,
respectively.\footnote{Most authors use notation such as $Y_i(1)$ and
$Y_i(0)$, or $Y_{1i}$ and $Y_{0i}$, for potential outcomes.
Freedman's (\citeyear{Freed08b}) choice of $a_i$ and $b_i$ helps make
the finite-population asymptotics more readable.
}
Thus, the observed outcome is
$
Y_i = a_i T_i + b_i (1 - T_i)
$,
where $T_i$ is a dummy variable for treatment $A$.

Random assignment is the sole source of randomness in this
model.
The $n$ subjects are the population of interest; they
are not assumed to be randomly drawn from a superpopulation.
For each subject, $a_i$, $b_i$, and $\mb{z}_i$ are fixed,
but $T_i$ and thus~$Y_i$ are random.

Let $\ov{a}$, $\ov{a}_A$, and $\ov{a}_B$ denote
the means of $a_i$ over the population, treatment group~$A$, and
treatment group $B$:
\[
\ov{a} = \frac{1}{n} \sum_{i=1}^n
a_i,\qquad \ov{a}_A = \frac{1}{n_A} \sum
_{i \in A} a_i,\qquad \ov{a}_B =
\frac{1}{n - n_A} \sum_{i \in B} a_i.
\]
Use similar notation for the means of
$b_i$, $Y_i$, $\mb{z}_i$, and other variables.

Our goal is to estimate the average treatment effect of $A$ relative
to $B$:
\[
\operatorname{ATE} = \ov{a} - \ov{b}.
\]

%s2.2 #&#
\subsection{Estimators of average treatment effect}\label{sec2.2}
The unadjusted or difference-in-means estimator of $\operatorname{ATE}$ is
\[
\hatw{\operatorname{ATE}}_{\mathrm{unadj}} = \ov{Y}_A -
\ov{Y}_B = \ov {a}_A - \ov{b}_B.
\]

The usual OLS-adjusted estimator of $\operatorname{ATE}$ is
the estimated coefficient on~$T_i$ in the OLS regression of
$Y_i$ on $T_i$ and $\mb{z}_i$.
(All
regressions described in this paper include intercepts.)
Let $\hatw{\operatorname{ATE}}_{\mathrm{adj}}$ denote this estimator.

A third estimator, $\hatw{\operatorname{ATE}}_{\mathrm{interact}}$,
can be computed as the estimated coefficient on
$T_i$ in the OLS regression of
$Y_i$ on $T_i$, $\mb{z}_i$, and $T_i (\mb{z}_i - \ov{\mb{z}})$.
Section~\ref{sec3} motivates
this estimator by analogy with regression
estimators in survey sampling.
In the context of observational studies,
Imbens and Wooldridge [(\citeyear{ImbensWool}), pp. 28--30] give a
theoretical analysis of $\hatw{\operatorname{ATE}}_{\mathrm{interact}}$, and
a related method is known as the
Peters--Belson or Oaxaca--Blinder estimator.\footnote{See
\citet{Cochran69}, \citet{Rubin84}, and
\citet{Kline}. \citet{Hansen} analyze a randomized experiment
with a variant of the Peters--Belson estimator derived from logistic
regression.}
When $\mb{z}_i$ is a set of indicators
for the values of a categorical variable,
$\hatw{\operatorname{ATE}}_{\mathrm{interact}}$ is equivalent to
subclassification or
poststratification [\citet{Miratrix}].

%s3 #&#
\section{Connections with sampling}\label{sec3}
Cochran [(\citeyear{Cochran77}), Chapter 7] gives a very readable discussion of
regression estimators in sampling.\footnote{See also
Fuller (\citeyear{Fuller02}, \citeyear{Fuller09}).}
In one example [\citet{Watson}], the goal was to
estimate $\ov{y}$, the average surface area of the leaves on a plant.
Measuring a leaf's area is time-consuming,
but its weight can be found quickly.
So the researcher weighed all the leaves, but
measured area for only a small sample. In simple random
sampling, the sample mean area $\ov{y}_S$
is an unbiased estimator of $\ov{y}$.
But $\ov{y}_S$ ignores the auxiliary data on leaf weights.
The sample and population mean weights ($\ov{z}_S$ and
$\ov{z}$) are
both known, and if $\ov{z} > \ov{z}_S$, then we expect that
$\ov{y} > \ov{y}_S$.
This motivates a ``linear regression estimator''
%e1 #&#
%
\begin{equation}
\hatw{\ov{y}}_{\mathrm{reg}} = \ov{y}_S + q (\ov{z} -
\ov{z}_S), \label{reg}
\end{equation}
where $q$ is an adjustment factor.
One way to choose $q$ is to
regress leaf area on leaf weight in the sample.

Regression adjustment in randomized experiments can be motivated
analogously
under the Neyman model.
The potential outcome $a_i$ is measured for
only a simple random sample (treatment group $A$), but
the covariates $\mb{z}_i$ are measured
for the whole population (the $n$ subjects). The sample mean
$\ov{a}_A$ is an unbiased estimator of $\ov{a}$, but it
ignores the auxiliary data on $\mb{z}_i$.
If the covariates are of some help in predicting $a_i$, then
another estimator to consider is
%e2 #&#
%
\begin{equation}
\hatw{\ov{a}}_{\mathrm{reg}} = \ov{a}_A + (\ov{\mb{z}} - \ov{
\mb{z}}_A) \mb{q}_a, \label{rega}
\end{equation}
where $\mb{q}_a$ is a $K \times1$ vector of
adjustment factors.
Similarly, we can consider using
%e3 #&#
%
\begin{equation}
\hatw{\ov{b}}_{\mathrm{reg}} = \ov{b}_B + (\ov{\mb{z}} - \ov{
\mb{z}}_B) \mb{q}_b \label{regb}
\end{equation}
to estimate $\ov{b}$ and then
$\hatw{\ov{a}}_{\mathrm{reg}} - \hatw{\ov{b}}_{\mathrm{reg}}$
to estimate $\operatorname{ATE} = \ov{a} - \ov{b}$.

The analogy suggests deriving
$\mb{q}_a$ and $\mb{q}_b$ from OLS regressions of
$a_i$ on $\mb{z}_i$ in treatment group $A$
and $b_i$ on $\mb{z}_i$ in treatment group $B$---in other words,\vspace*{1pt}
separate regressions of $Y_i$ on $\mb{z}_i$ in the two treatment groups.
The estimator $\hatw{\ov{a}}_{\mathrm{reg}} - \hatw{\ov
{b}}_{\mathrm{reg}}$
is then just $\hatw{\operatorname{ATE}}_{\mathrm{interact}}$.
If, instead, we use a pooled regression of
$Y_i$ on $T_i$ and $\mb{z}_i$ to derive a single vector
$\mb{q}_a = \mb{q}_b$,
then we get $\hatw{\operatorname{ATE}}_{\mathrm{adj}}$.

Connections between regression adjustment in experiments and
regression estimators in sampling have been noted
but remain underexplored.\footnote{Connections are noted by
\citet{Fienberg},
\citet{Hansen},
and \citet{Middleton}
but are not mentioned by
Cochran
despite his important contributions
to both literatures. He takes a design-based (agnostic) approach in
much of his work on sampling, but assumes a regression model in his
classic overview of regression adjustment in experiments
and observational studies [\citet{Cochran57}].
}
All three of the issues that Freedman\vadjust{\goodbreak} raised have parallels in the
sampling literature.
Under simple random sampling, when the
regression model is incorrect,
OLS adjustment of the estimated mean
still improves or does not hurt asymptotic precision [\citet{Cochran77}],
consistent
standard error estimators are available [\citet{Fuller75}], and the
adjusted estimator of the mean
has a small-sample bias [\citet{Cochran42}].

%s4 #&#
\section{Asymptotic precision}\label{sec4}

%s4.1 #&#
\subsection{Precision improvement in sampling}

This subsection gives an informal argument, adapted from \citet{Cochran77},
to show that in simple random sampling, OLS adjustment of the
sample mean improves or does not hurt asymptotic precision, even when
the regression model is incorrect.
Regularity conditions and other technical details are
omitted; the purpose is to motivate the
results on completely randomized experiments in
Section~\ref{sec4.2}.

First imagine using a ``fixed-slope'' regression estimator,
where $q$ in equation~(\ref{reg}) is fixed at
some value $q_0$ before sampling:
\[
\hatw{\ov{y}}_f = \ov{y}_S + q_0 (\ov{z}
- \ov{z}_S).
\]
If $q_0 = 0$, $\hatw{\ov{y}}_f$ is just $\ov{y}_S$.
More generally, $\hatw{\ov{y}}_f$ is the sample mean of
$
y_i - q_0 (z_i - \ov{z})
$,
so its variance follows the usual formula
with a finite-population correction:
\[
\var ( \hatw{\ov{y}}_f ) = \frac{N-n}{N-1} \frac{1}{n}
\frac{1}{N} \sum_{i=1}^N
\bigl[(y_i - \ov{y}) - q_0(z_i - \ov{z})
\bigr]^2,
\]
where $N$ is the population size and $n$ is the sample size.

Thus, choosing $q_0$ to minimize the variance of $\hatw{\ov{y}}_f$
is equivalent to running an OLS regression of
$y_i$ on $z_i$ in the
population. The solution is the ``population least squares'' slope,
\[
q_{\mathrm{PLS}} = \frac{
\sum_{i=1}^N (z_i - \ov{z})(y_i - \ov{y})
}{
\sum_{i=1}^N (z_i - \ov{z})^2
},
\]
and the minimum-variance fixed-slope regression estimator is
\[
\hatw{\ov{y}}_{\mathrm{PLS}} = \ov{y}_S + q_{\mathrm{PLS}} (\ov {z}
- \ov{z}_S).
\]
Since the sample mean $\ov{y}_S$ is a fixed-slope regression
estimator, it follows
that $\hatw{\ov{y}}_{\mathrm{PLS}}$
has lower variance than the sample mean, unless
$q_{\mathrm{PLS}} = 0$ (in which case $\hatw{\ov{y}}_{\mathrm{PLS}}
= \ov{y}_S$).

The actual OLS regression estimator
is almost as precise as $\hatw{\ov{y}}_{\mathrm{PLS}}$ in
sufficiently large
samples.
The difference between the two estimators is
\[
\hatw{\ov{y}}_{\mathrm{OLS}} - \hatw{\ov{y}}_{\mathrm{PLS}} = (
\hatw{q}_{\mathrm{OLS}} - q_{\mathrm{PLS}}) (\ov{z} - \ov{z}_S),
\]
where $\hatw{q}_{\mathrm{OLS}}$ is the estimated slope from a
regression of
$y_i$ on $z_i$ in the sample.
The estimation errors\vadjust{\goodbreak}
$\hatw{q}_{\mathrm{OLS}} - q_{\mathrm{PLS}}$, $\ov{z}_S - \ov{z}$, and
$\hatw{\ov{y}}_{\mathrm{PLS}} - \ov{y}$ are of order $1/\sqrt{n}$ in
probability.
Thus, the difference $\hatw{\ov{y}}_{\mathrm{OLS}} - \hatw{\ov
{y}}_{\mathrm{PLS}}$
is of order $1/n$, which is negligible
compared to the estimation error in $\hatw{\ov{y}}_{\mathrm{PLS}}$
when $n$ is large enough.

In sum, in large enough samples,
\[
\var (\hatw{\ov{y}}_{\mathrm{OLS}} ) \approx \var (\hatw{\ov{y}}_{\mathrm{PLS}}
) \leq \var (\ov{y}_S )
\]
and the inequality is strict unless $y_i$ and $z_i$ are
uncorrelated in the \mbox{population}.

%s4.2 #&#
\subsection{Precision improvement in experiments}\label{sec4.2}

The sampling result naturally leads to the conjecture that in a
completely randomized experiment, OLS adjustment with a full set of
treatment $\times$ covariate interactions improves or does not hurt
asymptotic precision,\vspace*{1pt} even when the regression model is incorrect. The
adjusted estimator $\hatw{\operatorname{ATE}}_{\mathrm{interact}}$ is just
the difference
between two OLS regression estimators from sampling theory,
while $\hatw{\operatorname{ATE}}_{\mathrm{unadj}}$ is the difference
between two sample means.

The conjecture is confirmed below. To summarize the results:
\begin{longlist}
\item[(1)]$\hatw{\operatorname{ATE}}_{\mathrm{interact}}$ is consistent and
asymptotically
normal (as are $\hatw{\operatorname{ATE}}_{\mathrm{unadj}}$ and $\hatw
{\operatorname{ATE}}_{\mathrm{adj}}$, from
Freedman's results).
\item[(2)] Asymptotically, $\hatw{\operatorname{ATE}}_{\mathrm{interact}}$ is
at least as efficient
as $\hatw{\operatorname{ATE}}_{\mathrm{unadj}}$, and more efficient unless
the covariates
are uncorrelated with the weighted average
\[
\frac{n-n_A}{n}a_i + \frac{n_A}{n}b_i.
\]
\item[(3)] Asymptotically, $\hatw{\operatorname{ATE}}_{\mathrm{interact}}$ is
at least as efficient
as $\hatw{\operatorname{ATE}}_{\mathrm{adj}}$, and more efficient unless
(a) the two
treatment groups have equal size or (b)
the covariates are uncorrelated with the treatment effect
$a_i - b_i$.
\end{longlist}

%s4.2.1 #&#
\subsubsection{Assumptions for asymptotics}
Finite-population asymptotic results are statements about
randomized experiments on (or random samples from) an
imaginary
infinite sequence of finite populations,
with increasing $n$.
The regularity conditions (assumptions on the limiting behavior of
the sequence) may seem vacuous, since one can
always construct a sequence that contains the actual
population and still satisfies the conditions.
But it may be useful to ask whether a sequence that preserves
any relevant ``irregularities''
(such as the influence of gross outliers) would violate the regularity
conditions.
See also Lumley [(\citeyear{Lumley}), pp. 217--218].

The asymptotic results in this paper assume Freedman's (\citeyear{Freed08b})
regularity conditions, generalized to allow multiple covariates;
the number of covariates $K$ is constant as $n$ grows.
One practical interpretation of these conditions is that in order for
the results to be applicable, the size of each treatment group should
be sufficiently large (and much larger than the number of covariates),
the influence of outliers should be small,
and near-collinearity in the covariates should be avoided.\vadjust{\goodbreak}

As \citet{Freed08a} notes, in principle, there should be an extra
subscript to index the sequence of populations:
for example, in the population with $n$ subjects,
the $i$th subject has potential outcomes $a_{i,n}$
and $b_{i,n}$, and the average treatment effect is $\operatorname{ATE}_{n}$.
Like Freedman, I drop the extra subscripts.

%co1 #&#
%
\begin{cond}\label{co1}
There is a bound
$L < \infty$ such that for all $n = 1, 2, \ldots$ and
$k = 1, \ldots, K$,
\[
\frac{1}{n} \sum_{i=1}^n
a_i^4 < L,\qquad \frac{1}{n} \sum
_{i=1}^n b_i^4 < L,\qquad
\frac{1}{n} \sum_{i=1}^n
z_{ik}^4 < L.
\]
\end{cond}

%co2 #&#
%
\begin{cond}\label{co2}
Let $\mb{Z}$ be the $n \times(K + 1)$ matrix whose $i$th row is
$(1, \mb{z}_i)$.
Then $n^{-1} \mb{Z}'\mb{Z}$ converges to a finite, invertible
matrix.
Also,
the population means of $a_i$, $b_i$, $a_i^2$, $b_i^2$,
$a_ib_i$, $a_i \mb{z}_i$, and $b_i \mb{z}_i$
converge to finite limits. For example,
$\lim_{n \goto\infty} n^{-1} \sum_{i=1}^n a_i \mb{z}_i$
exists and is a finite vector.
\end{cond}

%co3 #&#
%
\begin{cond}\label{co3}
The proportion $n_A/n$ converges to a limit $p_A$, with $0 <  p_A < 1$.
\end{cond}

%s4.2.2 #&#
\subsubsection{Asymptotic results}\label{sec4.2.2}
Let $\mb{Q}_a$ denote the limit of the vector of slope coefficients
in the population least squares regression of $a_i$ on
$\mb{z}_i$, that is,
\[
\mb{Q}_a = \lim_{n \goto\infty} \Biggl[ \Biggl(\sum
_{i=1}^n (\mb{z}_i - \ov{
\mb{z}})' (\mb{z}_i - \ov{\mb {z}})
\Biggr)^{-1} \sum_{i=1}^n (
\mb{z}_i - \ov{\mb{z}})' (a_i - \ov{a})
\Biggr].
\]
Define $\mb{Q}_b$ analogously.

Now define the prediction errors
\[
a^{*}_i = (a_i - \ov{a}) - (
\mb{z}_i - \ov{\mb{z}}) \mb{Q}_a ,\qquad b^{*}_i
= (b_i - \ov{b}) - (\mb{z}_i - \ov{\mb{z}})
\mb{Q}_b
\]
for $i = 1, \ldots, n$.

For any variables $x_i$ and $y_i$, let $\sigma^2_x$ and $\sigma_{x,y}$
denote the population variance of $x_i$ and the population covariance
of $x_i$ and $y_i$. For example,
\[
\sigma_{a^{*}, b^{*}} = \frac{1}{n} \sum_{i=1}^n
\bigl( a^{*}_i - \ov{a^{*}} \bigr) \bigl(
b^{*}_i - \ov{b^{*}} \bigr) = \frac{1}{n}
\sum_{i=1}^n a^{*}_i
b^{*}_i.
\]

Theorem~\ref{th1} and its corollaries are proved in the supplementary material [\citet{Lin}].

%th1 #&#
%
\begin{thmm}\label{th1}
Assume Conditions~\ref{co1}--\ref{co3}. Then
$\sqrt{n} (\hatw{\operatorname{ATE}}_{\mathrm{interact}} - \operatorname{ATE})$
converges in distribution to a Gaussian random variable with mean 0
and variance
\[
\frac{1-p_A}{p_A} \lim_{n \goto\infty} \sigma^2_{a^{*}} +
\frac{p_A}{1-p_A} \lim_{n \goto\infty} \sigma^2_{b^{*}} + 2
\lim_{n \goto\infty} \sigma_{a^{*}, b^{*}}.\vadjust{\goodbreak}
\]
\end{thmm}

%co1.1 #&#
%
\begin{cor}\label{cor1.1}
Assume Conditions~\ref{co1}--\ref{co3}.
Then $\hatw{\operatorname{ATE}}_{\mathrm{unadj}}$ has at least as much
asymptotic variance as
$\hatw{\operatorname{ATE}}_{\mathrm{interact}}$.
The difference is
\[
\frac{1}{np_A(1-p_A)} \lim_{n \goto\infty} \sigma^2_E,
\]
where $E_i = (\mb{z}_i - \ov{\mb{z}}) \mb{Q}_E$ and
$
\mb{Q}_E = (1 - p_A) \mb{Q}_a + p_A \mb{Q}_b
$.
Therefore, adjustment with $\hatw{\operatorname{ATE}}_{\mathrm{interact}}$
helps asymptotic precision if
$\mb{Q}_E \neq\mb{0}$ and is neutral if \mbox{$\mb{Q}_E = \mb{0}$}.
\end{cor}

%re1 #&#
%
\begin{com*}
(i)
$\mb{Q}_E$ can be thought of as a weighted average of $\mb{Q}_a$ and
$\mb{Q}_b$,
or as the limit of the vector of slope coefficients in the population
least squares regression of
$
(1 - p_A) a_i + p_A b_i
$
on $\mb{z}_i$.\vspace*{-6pt}
\begin{longlist}[(iii)]
\item[(ii)]
The weights may seem counterintuitive at first, but the sampling
analogy and equations (\ref{rega}) and (\ref{regb}) can help. Other things
being equal, adjustment has a larger effect on the estimated mean
from the smaller treatment group, because its mean covariate values
are further away from the population mean.
The adjustment added to $\ov{a}_A$ is
\[
(\ov{\mb{z}} - \ov{\mb{z}}_A) \hatw{\mb{Q}}_a =
\frac{n - n_A}{n}(\ov{\mb{z}}_B - \ov{\mb{z}}_A) \hatw{
\mb{Q}}_a,
\]
while the adjustment added to $\ov{b}_B$ is
\[
(\ov{\mb{z}} - \ov{\mb{z}}_B) \hatw{\mb{Q}}_b = -
\frac{n_A}{n}(\ov{\mb{z}}_B - \ov{\mb{z}}_A) \hatw{
\mb{Q}}_b,
\]
where $\hatw{\mb{Q}}_a$ and $\hatw{\mb{Q}}_b$ are OLS estimates that
converge to $\mb{Q}_a$ and $\mb{Q}_b$.

\item[(iii)]
If the covariates' associations with $a_i$ and $b_i$ go in opposite
directions, it is possible for adjustment with $\hatw{\operatorname{ATE}}_{\mathrm{interact}}$
to have no effect on asymptotic precision. Specifically,
if
$
(1 - p_A) \mb{Q}_a = -p_A \mb{Q}_b
$,
the adjustments to $\ov{a}_A$ and $\ov{b}_B$ tend to cancel each
other out.

\item[(iv)]
In designs with more than two treatment groups, estimators analogous
to $\hatw{\operatorname{ATE}}_{\mathrm{interact}}$ can be derived from a separate
regression in each treatment group, or, equivalently, a single
regression with the appropriate treatment dummies, covariates, and
interactions. The resulting estimator of (e.g.) $\ov{a} -
\ov{b}$ is at least as efficient as $\ov{Y}_A - \ov{Y}_B$, and more
efficient unless the covariates are uncorrelated with both $a_i$ and
$b_i$. The supplementary material [\citet{Lin}] gives a proof.
\end{longlist}
\end{com*}

%co1.2 #&#
%
\begin{cor}\label{cor1.2}
Assume Conditions~\ref{co1}--\ref{co3}.
Then $\hatw{\operatorname{ATE}}_{\mathrm{adj}}$ has at least as much
asymptotic variance as
$\hatw{\operatorname{ATE}}_{\mathrm{interact}}$.
The difference is
\[
\frac{(2 p_A - 1)^2}{np_A(1-p_A)} \lim_{n \goto\infty} \sigma^2_D,
\]
where $D_i = (\mb{z}_i - \ov{\mb{z}}) (\mb{Q}_a - \mb{Q}_b)$.
Therefore, the two estimators have equal asymptotic precision if
$p_A = 1/2$ or $\mb{Q}_a = \mb{Q}_b$. Otherwise,
$\hatw{\operatorname{ATE}}_{\mathrm{interact}}$ is asymptotically more efficient.
\end{cor}

%re2 #&#
%
\begin{com*}
(i)
$\mb{Q}_a - \mb{Q}_b$ is the limit of the vector of slope coefficients
in the population least squares regression of the treatment effect
$a_i - b_i$ on $\mb{z}_i$.\vspace*{-6pt}
\begin{longlist}[(iii)]
\item[(ii)]
For intuition about the behavior of $\hatw{\operatorname{ATE}}_{\mathrm{adj}}$,
suppose there is a single covariate, $z_i$, and the population least
squares slopes are $Q_a = 10$ and $Q_b = 2$.
Let~$\hatw{Q}$ denote the estimated coefficient on $z_i$ from
a pooled OLS regression of $Y_i$ on $T_i$ and $z_i$.
In sufficiently large samples, $\hatw{Q}$ tends to fall close to
$p_A Q_a + (1-p_A) Q_b$. Consider two cases:
\begin{itemize}
\item If the two treatment groups have equal size, then
$\ov{z} - \ov{z}_B = - (\ov{z} - \ov{z}_A)$, so
when $\ov{z} - \ov{z}_A = 1$,
the ideal linear adjustment
would add 10 to $\ov{a}_A$ and subtract 2 from $\ov{b}_B$.
Instead, $\hatw{\operatorname{ATE}}_{\mathrm{adj}}$ uses the pooled slope estimate
$\hatw{Q} \approx6$, so it
tends to underadjust $\ov{a}_A$ (adding
about 6) and
overadjust $\ov{b}_B$ (subtracting about 6).
Two wrongs make a right:
the adjustment adds about 12 to
$\ov{a}_A - \ov{b}_B$, just as
$\hatw{\operatorname{ATE}}_{\mathrm{interact}}$ would have done.

\item If group $A$ is 9 times larger than group $B$,
then $\ov{z} - \ov{z}_B = - 9(\ov{z} - \ov{z}_A)$, so
when $\ov{z} - \ov{z}_A = 1$,
the ideal linear adjustment
adds 10 to $\ov{a}_A$ and subtracts
$9 \cdot2 = 18$ from $\ov{b}_B$, thus adding 28 to the
estimate of $\operatorname{ATE}$.
In contrast, the pooled adjustment adds $\hatw{Q} \approx9.2$
to $\ov{a}_A$ and subtracts
$9 \hatw{Q} \approx82.8$ from $\ov{b}_B$, thus adding
about 92 to the estimate of $\operatorname{ATE}$. The problem is that the pooled
regression has more observations of $a_i$ than of $b_i$, but the
adjustment has a larger effect on the estimate of $\ov{b}$
than on that of $\ov{a}$, since
group $B$'s mean covariate value is further away from the population
mean.
\end{itemize}

\item[(iii)]
The example above suggests an alternative regression adjustment:
when group $A$ has nine-tenths of the subjects, give group $B$
nine-tenths of the weight.
More generally, let $\tilde{p}_A = n_A/n$.
Run a weighted least squares regression of $Y_i$ on~$T_i$ and
$\mb{z}_i$, with weights of $(1 - \tilde{p}_A)/\tilde{p}_A$ on each
observation from group $A$ and $\tilde{p}_A/(1-\tilde{p}_A)$ on each
observation from group $B$.
This ``tyranny of the minority'' estimator is asymptotically
equivalent to $\hatw{\operatorname{ATE}}_{\mathrm{interact}}$ (the
supplementary material [\citet{Lin}] outlines a proof).
It is equal to
$\hatw{\operatorname{ATE}}_{\mathrm{adj}}$ when $\tilde{p}_A = 1/2$.

\item[(iv)]
The tyranny estimator can also be seen as a one-step variant of
Rubin and van der Laan's (\citeyear{Rubin11}) two-step ``targeted ANCOVA.''
Their estimator is equivalent to the difference in means of the
residuals from a
weighted least squares regression of $Y_i$ on $\mb{z}_i$, with the
same weights as in remark (iii).

\item[(v)]
When is the usual adjustment worse than no adjustment?
Equation (23) in \citet{Freed08a} implies that
with a single covariate $z_i$,
for
$\hatw{\operatorname{ATE}}_{\mathrm{adj}}$ to have higher asymptotic
variance than
$\hatw{\operatorname{ATE}}_{\mathrm{unadj}}$, a necessary (but not
sufficient) condition is
that
either the design must be so imbalanced that
more than three-quarters of the subjects are assigned to one group,
or $z_i$ must have a
larger covariance with the treatment effect $a_i - b_i$
than with the expected outcome $p_A a_i + (1-p_A)b_i$.
With multiple covariates, a similar condition can be derived from
equation (14) in \citet{Schochet}.

\item[(vi)]
With more than two treatment groups,
the usual adjustment can be
worse than no adjustment even when the design is balanced
[\citet{Freed08b}].
All the groups are pooled in a single
regression without treatment $\times$ covariate interactions,
so group $B$'s data can affect the contrast between $A$ and
$C$.
\end{longlist}
\end{com*}

%s4.2.3 #&#
\subsubsection{Example}
This simulation illustrates some of the key ideas.
\begin{longlist}[(1)]
\item[(1)] For $n = 1000$ subjects, a covariate $z_i$ was drawn
from the uniform distribution on $[-4, 4]$.
The potential outcomes were then generated as
\begin{eqnarray*}
a_{i} &=& \frac{\exp(z_i) + \exp(z_i/2)}{4} + \nu_i,
\\
b_{i} &=& \frac{-\exp(z_i) + \exp(z_i/2)}{4} + \varepsilon_i,
\end{eqnarray*}
with $\nu_i$ and $\varepsilon_i$ drawn independently from the
standard normal distribution.

\item[(2)] A completely randomized experiment was simulated 40,000 times,
assigning $n_A = 750$ subjects to treatment $A$
and the remainder to treatment~$B$.

\item[(3)] Step 2 was repeated for four other values of $n_A$ (600, 500,
400, and~250).
\end{longlist}

These are adverse conditions for regression adjustment:
$z_i$ covaries much more with the treatment effect $a_i -
b_i$ than with the potential outcomes, and
the population least squares slopes $Q_a = 1.06$ and $Q_b = -0.73$ are
of opposite signs.

Table~\ref{sim} compares $\hatw{\operatorname{ATE}}_{\mathrm{unadj}}$,
$\hatw{\operatorname{ATE}}_{\mathrm{adj}}$, $\hatw{\operatorname{ATE}}_{\mathrm{interact}}$,
and the ``tyranny of the minority'' estimator from remark (iii)
after Corollary~\ref{cor1.2}.
The first panel shows the asymptotic standard errors derived from
Freedman's (\citeyear{Freed08b}) Theorems 1 and 2
and this paper's Theorem~\ref{th1}
(with limits replaced by actual population values).
The second and third panels show the empirical standard deviations and
bias estimates
from the Monte Carlo simulation.

%t1 #&#
%
\begin{table}
\tabcolsep=0pt
\caption{Simulation (1000 subjects; 40,000 replications)}
\label{sim}
\begin{tabular*}{\textwidth}{@{\extracolsep{4in minus
4in}}ld{3.0}d{2.0}d{2.0}d{2.0}d{3.0}@{}}
\hline
& \multicolumn{5}{c@{}}{\textbf{Proportion assigned to treatment}
$\bolds{A}$}
\\[-4pt]
& \multicolumn{5}{c@{}}{\hrulefill}\\
\multicolumn{1}{@{}l}{\textbf{Estimator}}& \multicolumn
{1}{c}{$\bolds{0.75}$} & \multicolumn{1}{c}{$\bolds{0.6}$} &
\multicolumn{1}{c}{$\bolds{0.5}$} &
\multicolumn{1}{c}{$\bolds{0.4}$} & \multicolumn{1}{c@{}}{$\bolds{0.25}$}
\\
\hline
\multicolumn{6}{@{}l}{\textit{SD (asymptotic) $\times$ 1000}} \\
Unadjusted & 93 & 49 & 52 & 78 & 143
\\
Usual OLS-adjusted & 171 & 72 & 46 & 79 & 180
\\
OLS with interaction & 80 & 49 & 46 & 58 & 98
\\
Tyranny of the minority & 80 & 49 & 46 & 58 & 98
\\[3pt]
\multicolumn{6}{@{}l}{\textit{SD (empirical) $\times$ 1000}}
\\
Unadjusted & 93 & 49 & 53 & 78 & 142
\\
Usual OLS-adjusted & 171 & 73 & 47 & 80 & 180
\\
OLS with interaction & 81 & 50 & 47 & 59 & 99
\\
Tyranny of the minority & 81 & 50 & 47 & 59 & 99
\\[3pt]
\multicolumn{6}{@{}l}{\textit{Bias (estimated) $\times$ 1000}}
\\
Unadjusted & 0 & 0 & 0 & 0 & -2
\\
Usual OLS-adjusted & -3 & -3 & -3 & -3 & -5
\\
OLS with interaction & -5 & -3 & -3 & -4 & -6
\\
Tyranny of the minority & -5 & -3 & -3 & -4 & -6
\\
\hline
\end{tabular*}
\end{table}

The empirical standard deviations are very close to the asymptotic
predictions, and the estimated biases are small in comparison.
The usual adjustment hurts precision except when $n_A/n =
0.5$. In contrast,
$\hatw{\operatorname{ATE}}_{\mathrm{interact}}$ and the tyranny estimator
improve precision
except when $n_A/n = 0.6$. [This is approximately the value of~$p_A$
where $\hatw{\operatorname{ATE}}_{\mathrm{interact}}$ and $\hatw{\operatorname{ATE}}_{\mathrm{unadj}}$ have equal
asymptotic variance; see remark~(iii) after Corollary~\ref{cor1.1}.]

Randomization does not ``justify''
the regression model of
$\hatw{\operatorname{ATE}}_{\mathrm{interact}}$, and the linearity
assumption is far from
accurate in this example, but the estimator solves Freedman's
asymptotic precision problem.

%s5 #&#
\section{Variance estimation}\label{sec5}

Eicker (\citeyear{Eicker}) and White (\citeyear{White80a}, \citeyear{White80b}) proposed a
covariance matrix estimator for OLS that is
consistent under simple random sampling from an infinite population.
The regression model assumptions, such as linearity and
homoskedasticity, are not needed for this
result.\footnote{See, for example, Chamberlain [(\citeyear{Chamberlain}), pp. 17--19] or
Angrist and Pischke [(\citeyear{MHE}), pp. 40--48].
\citet{Fuller75} proves a finite-population version of the result.
}
The estimator is
\[
\bigl(\mb{X}'\mb{X}\bigr)^{-1} \mb{X}'
\mbox{diag}\bigl(\hat{\varepsilon}_1^2, \ldots, \hat{
\varepsilon}_n^2\bigr) \mb{X} \bigl(\mb{X}'
\mb{X}\bigr)^{-1},
\]
where
$\mb{X}$ is the matrix of regressors and
$\hat{\varepsilon}_i$ is the $i$th OLS residual.
It is known as the sandwich estimator because of its form, or as the
Huber--White estimator because it
is the sample analog of
Huber's (\citeyear{Huber}) formula for the asymptotic variance of
a maximum likelihood estimator when the model is incorrect.

Theorem~\ref{th2} shows that under the Neyman model,
the sandwich variance estimators for
$\hatw{\operatorname{ATE}}_{\mathrm{adj}}$ and $\hatw{\operatorname{ATE}}_{\mathrm{interact}}$ are consistent or
asymptotically conservative.
Together, Theorems~\ref{th1} and~\ref{th2} in this paper
and Theorem 2 in \citet{Freed08b}
imply that asymptotically valid\vspace*{2pt}
confidence intervals for $\operatorname{ATE}$ can be constructed from either
$\hatw{\operatorname{ATE}}_{\mathrm{adj}}$ or $\hatw{\operatorname{ATE}}_{\mathrm{interact}}$ and the sandwich
standard error estimator.

The vectors $\mb{Q}_a$ and $\mb{Q}_b$ were defined in Section~\ref{sec4.2.2}.
Let $\mb{Q}$ denote the weighted average
$p_A \mb{Q}_a + (1-p_A) \mb{Q}_b$.
As shown in \citet{Freed08b} and the supplementary material [\citet{Lin}],
$\mb{Q}$ is the probability limit of the vector of estimated
coefficients on $\mb{z}_i$ in the OLS regression of $Y_i$ on $T_i$ and~$\mb{z}_i$.

Mimicking Section~\ref{sec4.2.2}, define the prediction errors
\[
a^{**}_i = (a_i - \ov{a}) - (
\mb{z}_i - \ov{\mb{z}}) \mb{Q} ,\qquad b^{**}_i =
(b_i - \ov{b}) - (\mb{z}_i - \ov{\mb{z}}) \mb{Q}
\]
for $i = 1, \ldots, n$.

Theorem~\ref{th2} is proved in the supplementary material [\citet{Lin}].

%th2 #&#
%
\begin{thmm}\label{th2}
Assume Conditions~\ref{co1}--\ref{co3}.
Let $\hatw{v}_{\mathrm{adj}}$ and $\hatw{v}_{\mathrm{interact}}$
denote the sandwich
variance estimators for
$\hatw{\operatorname{ATE}}_{\mathrm{adj}}$ and $\hatw{\operatorname{ATE}}_{\mathrm{interact}}$.
Then $n \hatw{v}_{\mathrm{adj}}$ converges in probability to
\[
\frac{1}{p_A} \lim_{n \goto\infty} \sigma^2_{a^{**}} +
\frac{1}{1 - p_A} \lim_{n \goto\infty} \sigma^2_{b^{**}},
\]
which is greater than or equal to the true asymptotic variance of
$\sqrt{n} (\hatw{\operatorname{ATE}}_{\mathrm{adj}} - \operatorname{ATE})$.
The difference is
\[
\lim_{n \goto\infty} \sigma^2_{(a - b)} = \lim_{n \goto\infty}
\frac{1}{n} \sum_{i=1}^n
\bigl[(a_i - b_i) - \operatorname{ATE}
\bigr]^2.
\]
Similarly, $n \hatw{v}_{\mathrm{interact}}$ converges in probability to
\[
\frac{1}{p_A} \lim_{n \goto\infty} \sigma^2_{a^{*}} +
\frac{1}{1 - p_A} \lim_{n \goto\infty} \sigma^2_{b^{*}},
\]
which is greater than or equal to the true asymptotic variance of\break
$\sqrt{n} (\hatw{\operatorname{ATE}}_{\mathrm{interact}} - \operatorname{ATE})$. The difference is
\[
\lim_{n \goto\infty} \sigma^2_{(a^{*} - b^{*})} = \lim_{n \goto\infty}
\frac{1}{n} \sum_{i=1}^n
\bigl[(a_i - b_i) - \operatorname{ATE} - (
\mb{z}_i - \ov{\mb{z}}) (\mb{Q}_a -
\mb{Q}_b)\bigr]^2.
\]
\end{thmm}

%re3 #&#
%
\begin{com*}
(i) Theorem~\ref{th2} generalizes to designs with more than two treatment
groups.\vspace*{-4pt}
\begin{longlist}[(viii)]
\item[(ii)]
With two treatment groups of equal size, the
conventional OLS variance estimator
for $\hatw{\operatorname{ATE}}_{\mathrm{adj}}$
is also consistent or
asymptotically conservative [\citet{Freed08a}].

\item[(iii)]
\citet{Freed08a} shows analogous results for variance estimators for
the difference in means; the issue there is whether to assume
$\sigma^2_a = \sigma^2_b$.
\citet{Reichardt} and Freedman, Pisani, and Purves
[(\citeyear{FPP}), pp. 508--511]
give helpful expositions of basic results under the Neyman
model. Related\vadjust{\goodbreak} issues appear in discussions of the two-sample problem
[Miller (\citeyear{Miller}), pp. 56--62, \citet{Stonehouse}] and
randomization tests [\citet{Gail}, Chung and Romano (\citeyear{Chung2011a}, \citeyear{Chung2011b})].

\item[(iv)]
With a small sample or points of high leverage,
the sandwich estimator can have substantial downward bias
and high variability. \citet{MacKinnon} discusses bias-corrected
sandwich estimators and improved confidence intervals based on the
wild bootstrap. See also \citet{Wu},
Tibshirani (\citeyear{Tibshirani}),
Angrist and Pischke [(\citeyear{MHE}), Chapter 8],
and \citet{KlineSantos}.

\item[(v)]
When $\hatw{\operatorname{ATE}}_{\mathrm{unadj}}$ is computed by
regressing $Y_i$ on $T_i$,
the HC2 bias-corrected sandwich estimator
[\citet{MacWhite}, \citet{Royall},
Wu (\citeyear{Wu}), page~1274] gives
exactly
the variance estimate
preferred by \citet{Ney23} and \citet{Freed08a}:
$
\hatw{\sigma}^2_a /n_A + \hatw{\sigma}^2_b / (n - n_A)
$,
where $\hatw{\sigma}^2_a$ and $\hatw{\sigma}^2_b$
are the sample variances of $Y_i$ in the two groups.\footnote{For
details, see \citet{HinkleyWang},
Angrist and Pischke [(\citeyear{MHE}), pp.~294--304], or
\citet{Samii}.}

\item[(vi)]
When the $n$ subjects are randomly drawn from a superpopulation,
$\hatw{v}_{\mathrm{interact}}$ does not take into account
the variability in
$\ov{\mb{z}}$ [Imbens and Wooldridge (\citeyear{ImbensWool}), pp.~28--30].
In the Neyman model, $\ov{\mb{z}}$ is fixed.

\item[(vii)]
Freedman's (\citeyear{Freed06}) critique of the sandwich estimator does not apply
here, as $\hatw{\operatorname{ATE}}_{\mathrm{adj}}$ and $\hatw{\operatorname{ATE}}_{\mathrm{interact}}$ are consistent
even when their regression models are incorrect.

\item[(viii)]
\citet{Freed08a} associates the difference in means and regression
with heteroskedasticity-robust and conventional variance
estimators, respectively.
His rationale for these pairings is unclear.
The pooled-variance two-sample $t$-test and the conventional $F$-test
for equality
of means are often used in difference-in-means analyses.
Conversely, the sandwich estimator has become the usual variance
estimator for regression in economics [\citet{Stock}].
The question of whether to adjust for covariates should be
disentangled from the question of whether to assume homoskedasticity.
\end{longlist}
\end{com*}

%s6 #&#
\section{Bias}\label{sec6}
The bias of OLS adjustment diminishes rapidly with the number of
randomly assigned units:
$\hatw{\operatorname{ATE}}_{\mathrm{adj}}$ and $\hatw{\operatorname{ATE}}_{\mathrm{interact}}$ have biases of
order $1/n$, while their standard errors are of order
$1/\sqrt{n}$.
Brief remarks follow; see also Deaton [(\citeyear{Deaton}), pp. 443--444], Imbens
[(\citeyear{Imbens}), pp. 410--411], and \citet{Green}.
\begin{longlist}[(iii)]
\item[(i)]
If the actual random assignment yields
substantial covariate imbalance,
it is hardly reassuring to be told that the difference in means
is unbiased over all possible random
assignments. \citet{Senn} and
Cox and Reid [(\citeyear{CoxReid}), pp. 29--32]
argue that inference should be conditional on a measure of covariate
imbalance, and that the conditional bias of $\hatw{\operatorname{ATE}}_{\mathrm{unadj}}$
justifies adjustment.
\citet{Tukey91} suggests adjustment ``perhaps\vadjust{\goodbreak} as a supplemental
analysis'' for
``protection against either the consequences of inadequate
randomization or the (random) occurrence of an unusual
randomization.''

\item[(ii)]
As noted in Section~\ref{sec2.2}, poststratification is a special case of
$\hatw{\operatorname{ATE}}_{\mathrm{interact}}$.
The poststratified estimator is
a population-weighted average of
subgroup-specific differences in means.
Conditional on the numbers of
subgroup members assigned to each treatment, the poststratified
estimator is unbiased, but $\hatw{\operatorname{ATE}}_{\mathrm{unadj}}$
can be biased.
\citet{Miratrix} give finite-sample and
asymptotic analyses of poststratification and blocking;
see also \citet{Holt} in the sampling context.

\item[(iii)]
\citet{Cochran77} analyzes the bias of $\hatw{\ov{y}}_{\mathrm{reg}}$ in
equation (\ref{reg}). If the adjustment factor $q$ is fixed,
$\hatw{\ov{y}}_{\mathrm{reg}}$ is unbiased,
but if $q$ varies with the sample, $\hatw{\ov{y}}_{\mathrm{reg}}$
has a
bias of $- \cov(q, \ov{z}_S)$.
The leading term in the bias of $\hatw{\ov{y}}_{\mathrm{OLS}}$ is
\[
- \frac{1}{\sigma_z^2} \biggl(\frac{1}{n} - \frac{1}{N} \biggr)
\lim_{N \goto\infty} \frac{1}{N} \sum_{i=1}^N
e_i (z_i - \ov{z})^2,
\]
where $n$ is the sample size, $N$ is the population size,
and $e_i$ is the prediction error in the population least squares
regression of $y_i$ on $z_i$.

\item[(iv)]
By analogy, the leading term in the bias of
$\hatw{\operatorname{ATE}}_{\mathrm{interact}}$ (with a single covariate
$z_i$) is
\[
- \frac{1}{\sigma_z^2} \Biggl[ \biggl(\frac{1}{n_A} - \frac{1}{n}
\biggr) \lim_{n \goto\infty} \frac{1}{n} \sum_{i=1}^n
a_i^{*} (z_i - \ov{z})^2 -
\biggl(\frac{1}{n - n_A} - \frac{1}{n} \biggr) \lim_{n \goto\infty}
\frac{1}{n} \sum_{i=1}^n
b_i^{*} (z_i - \ov{z})^2 \Biggr].
\]
Thus, the bias tends to depend largely on $n$, $n_A/n$, and
the importance of
omitted quadratic terms in the regressions of
$a_i$ and $b_i$ on $z_i$. With multiple covariates, it would also
depend on the importance of omitted first-order interactions
between the covariates.

\item[(v)]
Remark (iii) also implies that if the adjustment factors
$\mb{q}_a$ and $\mb{q}_b$ in
equations (\ref{rega}) and (\ref{regb}) do not vary with
random assignment,
the resulting estimator of $\operatorname{ATE}$
is unbiased.
Middleton and Aronow's (\citeyear{Middleton}) insightful paper uses out-of-sample
data to determine $\mb{q}_a = \mb{q}_b$.
In-sample data can be used when
multiple pretests
(pre-randomization outcome measures) are available:
if the only covariate~$z_i$ is the most recent pretest,
a common adjustment factor $q_a = q_b$ can be determined by
regressing $z_i$ on an earlier pretest.\looseness=-1
\end{longlist}

%s7 #&#
\section{Empirical example}\label{sec7}
This section suggests empirical
checks on the asymptotic
approximations.
I will focus on the validity of confidence intervals, using data from
a social experiment for an illustrative example.

%s7.1 #&#
\subsection{Background}
Angrist, Lang, and Oreopoulos [(\citeyear{AngristLang}); henceforth ALO] conducted an
experiment to
estimate the effects of support services and financial incentives on
college students' academic achievement. At a Canadian university
campus, all first-year undergraduates entering in September 2005,
except those with a high-school grade point average (GPA) in the top
quartile, were randomly assigned to four groups.
One treatment group was offered support services
(peer advising and supplemental instruction).
Another group was offered financial incentives
(awards of \$1000 to \$5000 for meeting a target GPA).
A third group was offered both services and incentives.
The control group was eligible only for standard university support
services (which included supplemental instruction for some courses).

ALO report that for women, the
combination of services and incentives
had sizable estimated effects on both first- and
second-year academic achievement, even though the programs were only
offered during the first year. In contrast, there was no
evidence that services alone or incentives alone had
lasting effects for women or that any of the treatments improved
achievement for men
(who were much less likely to contact peer advisors).

To simplify the example
and focus on the accuracy of large-sample
approximations in samples that are not huge,
I use only the data for men (43 percent of the students) in the
services-and-incentives and services-only
groups (9~percent and 15 percent of the men).
First-year GPA data are available for
58 men in the services-and-incentives group and
99 in the services-only group.

Table~\ref{ests} shows alternative estimates of $\operatorname{ATE}$ (the average treatment
effect of the financial incentives, given that the support services
were available). The services-and-incentives and services-only groups
had average first-year GPAs of 1.82 and 1.86 (on a scale of 0 to 4),
so the unadjusted estimate of $\operatorname{ATE}$ is
close to zero. OLS adjustment for high-school GPA hardly makes a practical
difference to either the point estimate of $\operatorname{ATE}$ or the sandwich
standard error estimate, regardless of whether the treatment $\times$
covariate interaction is included.\footnote{ALO adjust
for a larger set of covariates, including first language, parents'
education, and self-reported procrastination tendencies. These also
have little effect on
the estimated standard errors.
}
The two groups had similar average
high-school GPAs,
and high-school GPA was not
a strong predictor of first-year college GPA.\looseness=-1

%t2 #&#
%
\begin{table}
\caption{Estimates of average treatment effect on men's first-year GPA}
\label{ests}
\begin{tabular*}{\textwidth}{@{\extracolsep{\fill}}lcc@{}}
\hline
& \textbf{Point estimate} & \textbf{Sandwich SE}\\
\hline
Unadjusted & $-0.036$ & 0.158
\\
Usual OLS-adjusted & $-0.083$ & 0.146
\\
OLS with interaction & $-0.081$ & 0.146
\\
\hline
\end{tabular*}
\end{table}

The finding that adjustment appears to have little effect on precision
is not unusual in social experiments,
because the covariates are often only weakly correlated with
the outcome\vadjust{\goodbreak}
[Meyer (\citeyear{Meyer}), pp. 100, 116, Lin
et al. (\citeyear{LinSSP}), pp. 129--133].
Examining eight social experiments with a wide range of outcome
variables, \citet{Schochet} finds
$R^2$ values above 0.3 only when the outcome is a
standardized achievement test score or Medicaid costs and the
covariates include a lagged outcome.

Researchers may prefer not to adjust when the expected precision
improvement is meager. Either way, confidence intervals for treatment
effects typically rely on
either strong parametric assumptions (such as a
constant treatment effect or a normally distributed outcome) or
asymptotic approximations. When a sandwich standard error estimate
is multiplied by 1.96 to form a margin of error for a 95 percent
confidence interval, the calculation assumes the sample is large
enough that (i) the estimator of $\operatorname{ATE}$ is approximately normally
distributed, (ii) the bias and variability of the sandwich standard
error estimator are small relative to the true standard error (or else
the bias is conservative and the variability is small), and
(iii) the bias of adjustment (if used) is small relative to the true
standard error.

Below I discuss a simulation to check for confidence interval
undercoverage due to violations of (i) or (ii), and a bias estimate
to check for violations of~(iii).
These checks are not foolproof, but
may provide a useful sniff test.

%s7.2 #&#
\subsection{Simulation}
For technical reasons, the most revealing initial check is a
simulation with a constant treatment effect.
When treatment effects are heterogeneous,
the sandwich standard error estimators for
$\hatw{\operatorname{ATE}}_{\mathrm{unadj}}$ and $\hatw{\operatorname{ATE}}_{\mathrm{adj}}$ are asymptotically
conservative,\footnote{By
Theorem~\ref{th2}, the
sandwich standard error estimator for $\hatw{\operatorname{ATE}}_{\mathrm
{interact}}$ is
also asymptotically conservative unless the treatment effect is a
linear function of the covariates.
}
so nominal 95 percent confidence intervals for $\operatorname{ATE}$ achieve
greater than 95 percent coverage in large enough samples.
A simulation that overstates treatment effect heterogeneity may
overestimate coverage.

%t3 #&#
%
\begin{table}
\def\arraystretch{0.9}
\caption{Simulation with zero treatment effect (250,000
replications). The fourth panel\break shows the empirical coverage rates
of nominal 95 percent confidence intervals.\break All other estimates are
on the four-point GPA scale}
\label{bench}
\begin{tabular*}{\textwidth}{@{\extracolsep{\fill}}ld{2.3}d{2.3}d{2.3}@{}}
\hline
& \multicolumn{3}{c@{}}{\textbf{ATE estimator}}
\\[-4pt]
& \multicolumn{3}{c@{}}{\hrulefill}\\
& \multicolumn{1}{c}{\textbf{Unadjusted}} & \multicolumn
{1}{c}{\textbf{Usual OLS-adjusted}} & \multicolumn{1}{c@{}}{\textbf
{OLS with interaction}}
\\
\hline
\textit{Bias \& SD of ATE estimator} & & &
\\
Mean (estimated bias) & 0.000 & 0.000 & 0.000
\\
SD & 0.158 & 0.147 & 0.147
\\[3pt]
\textit{Bias of SE estimator} & & &
\\
Classic sandwich & -0.001 & -0.002 & -0.002
\\
HC1 & 0.000 & 0.000 & 0.000
\\
HC2 & 0.000 & 0.000 & 0.000
\\
HC3 & 0.001 & 0.002 & 0.002
\\[3pt]
\textit{SD of SE estimator} & & &
\\
Classic sandwich & 0.004& 0.004&0.004
\\
HC1 & 0.004& 0.004&0.004
\\
HC2 & 0.004& 0.004&0.004
\\
HC3 & 0.004& 0.004&0.005
\\[3pt]
\textit{CI coverage (percent)} & & &
\\
Classic sandwich & 94.6 & 94.5 & 94.4
\\
HC1 & 94.8 & 94.7 & 94.7
\\
HC2 (normal) & 94.8 & 94.8 & 94.8
\\
HC2 (Welch $t$) & 95.1 & &
\\
HC3 & 95.0 & 95.0 & 95.1
\\[3pt]
\textit{CI width (average)} & & &
\\
Classic sandwich & 0.618 & 0.570 & 0.568
\\
HC1 & 0.622 & 0.576 & 0.575
\\
HC2 (normal) & 0.622 & 0.576 & 0.577
\\
HC2 (Welch $t$) & 0.629 & &
\\
HC3 & 0.627 & 0.583 & 0.586
\\
\hline
\end{tabular*}  \vspace*{-3pt}
\end{table}

Table~\ref{bench} reports a simulation that assumes
treatment had no effect on any of the men.
Keeping the GPA data at their actual values,
I replicated the experiment 250,000 times, each
time randomly assigning 58 men to services-and-incentives and 99 to
services-only. The first panel shows the means and standard deviations
of $\hatw{\operatorname{ATE}}_{\mathrm{unadj}}$, $\hatw{\operatorname{ATE}}_{\mathrm{adj}}$, and
$\hatw{\operatorname{ATE}}_{\mathrm{interact}}$. All three estimators are
approximately
unbiased, but adjustment slightly improves precision. Since the
simulation assumes a constant treatment effect (zero), including the
treatment $\times$ covariate interaction does not improve precision
relative to the usual adjustment.

The second and third panels show the estimated biases and standard
deviations of
the sandwich standard error estimator and the three variants discussed
in Angrist and Pischke
[(\citeyear{MHE}), pp. 294--308]. ALO's paper uses HC1 [\citet{Hinkley}], which
simply multiplies the
sandwich variance estimator\vadjust{\goodbreak} by $n / (n-k)$, where $k$ is the number of
regressors. HC2 [see remark~(v) after
Theorem~\ref{th2}] and the approximate jackknife HC3 [Davidson and Mac\-Kinnon
(\citeyear{DavidsonMac}), pages~553--554,
Tibshirani (\citeyear{Tibshirani})] inflate the squared residuals in
the sandwich
formula by the factors $(1-h_{ii})^{-1}$ and $(1-h_{ii})^{-2}$, where
$h_{ii}$ is the $i$th
diagonal element of the hat matrix $\mb{X} (\mb{X}'\mb{X})^{-1} \mb{X}'$.
All the standard error estimators appear to be approximately unbiased
with low variability.

The fourth and fifth panels evaluate thirteen ways of constructing a
95 percent
confidence interval. For each of the three estimators of $\operatorname{ATE}$, each
of the four standard error estimators was multiplied by 1.96 to form
the margin of error for a normal-approximation interval.
Welch's (\citeyear{Welch}) $t$-interval
[Miller (\citeyear{Miller}), pp.~60--62] was also
constructed. Welch's interval uses $\hatw{\operatorname{ATE}}_{\mathrm{unadj}}$,
the HC2 standard error estimator, and the $t$-distribution with the
Welch--Satterthwaite approximate degrees of freedom.

The fourth panel shows that all thirteen confidence intervals cover
the true value of $\operatorname{ATE}$ (zero) with approximately 95 percent
probability. The fifth panel shows the average widths of the
intervals. (The mean and median widths agree up to three decimal places.)
The regression-adjusted intervals are narrower on average than
the unadjusted intervals, but the improvement is meager.
In sum, adjustment appears to yield slightly more
precise inference without sacrificing validity.

%s7.3 #&#
\subsection{Bias estimates}
One limitation of the simulation above is that the bias of adjustment
may be larger when treatment effects are heterogeneous.
With a single covariate $z_i$, the leading term in the bias of
$\hatw{\operatorname{ATE}}_{\mathrm{adj}}$ is\footnote{An equivalent
expression appears in the
version of \citet{Freed08a} on his web page. It can be derived from
\citet{Freed08b} after correcting a minor error in equations (17) and (18):
the potential outcomes should be centered.
}
\[
- \frac{1}{n} \frac{1}{\sigma_z^2} \lim_{n \goto\infty} \frac{1}{n}
\sum_{i=1}^n \bigl[(a_i -
b_i) - \operatorname{ATE}\bigr] (z_i -
\ov{z})^2.
\]
Thus, with a constant treatment effect, the leading term is zero (and
the bias is of order $n^{-3/2}$ or smaller).
\citet{Freed08b} shows that with a balanced design and a constant
treatment effect, the bias is exactly zero.

We can estimate the leading term by rewriting it as
\[
- \frac{1}{n} \frac{1}{\sigma_z^2} \Biggl[ \lim_{n \goto\infty}
\frac{1}{n} \sum_{i=1}^n
(a_i - \ov{a}) (z_i - \ov{z})^2 -
\lim_{n \goto\infty} \frac{1}{n} \sum_{i=1}^n
(b_i - \ov{b}) (z_i - \ov{z})^2 \Biggr]
\]
and substituting the sample variance of high-school GPA for
$\sigma_z^2$,
and the sample covariances of first-year college GPA with
the square of centered high-school GPA in the services-and-incentives
and services-only groups for the bracketed limits.
The resulting estimate of the bias of $\hatw{\operatorname{ATE}}_{\mathrm{adj}}$
is $-0.0002$ on the four-point GPA scale.
Similarly, the leading term in the bias of $\hatw{\operatorname{ATE}}_{\mathrm{interact}}$
[Section~\ref{sec6}, remark (iv)] can be estimated, and the result is also
$-0.0002$. The biases would need to be orders of magnitude larger to
have noticeable effects on confidence interval coverage (the estimated
standard errors of $\hatw{\operatorname{ATE}}_{\mathrm{adj}}$
and $\hatw{\operatorname{ATE}}_{\mathrm{interact}}$ in Table~\ref{ests} are both 0.146).

%s7.4 #&#
\subsection{Remarks}
(i)
This exercise does not prove that the bias of adjustment is negligible,
since it
just replaces a first-order approximation (the bias is close
to zero in large enough samples) with a second-order approximation
(the bias is close to the leading term in large enough samples), and
the estimate of the leading term has sampling
error.\footnote{Finite-population bootstrap methods [Davison and
Hinkley (\citeyear{DavisonHinkley}), pp. 92--100, 125] may also be useful
for estimating the bias of $\hatw{\operatorname{ATE}}_{\mathrm
{interact}}$, but
similar caveats would apply.
}
The checks suggested here
cannot validate an analysis, but they can reveal problems.\vspace*{-6pt}
\begin{longlist}[(iii)]
\item[(ii)]
Another limitation is that the simulation assumes the potential
outcome distributions have the same shape. In Stonehouse and Forrester's (\citeyear{Stonehouse}) simulations, Welch's $t$-test was not robust to
extreme skewness in the smaller group when that group's sample size
was 30 or smaller. That does not appear to be a serious issue in
this example, however. The distribution of men's first-year GPA
in the services-and-incentives group is roughly symmetric (e.g., see
ALO, Figure 1A).

\item[(iii)]
The simulation check may appear to resemble permutation inference
[\citet{Fisher35}, \citet{Tukey93}, \citet{Rosenbaum02}], but the goals differ.
Here, the constant treatment effect scenario just gives
a benchmark to check the finite-sample coverage of confidence
intervals
that are asymptotically valid under weaker assumptions.
Classical permutation methods achieve exact inference under
strong assumptions about treatment effects,
but may give misleading results when the assumptions fail.
For example, the Fisher--Pitman
permutation test is asymptotically equivalent to a $t$-test using
the conventional OLS standard error estimator. The test can be inverted
to give
exact confidence intervals for a constant treatment effect, but these
intervals may undercover $\operatorname{ATE}$ when treatment effects are
heterogeneous and the design is imbalanced [\citet{Gail}].

\item[(iv)]
Chung and Romano (\citeyear{Chung2011a}, \citeyear{Chung2011b}) discuss and extend a literature on
permutation tests that do remain valid
asymptotically when the null hypothesis is weakened. One such test is
based on the permutation distribution of a heteroskedasticity-robust
$t$-statistic.
Exploration of this approach
under the Neyman model (with
and without covariate adjustment) would be valuable.
\end{longlist}

%s8 #&#
\section{Further remarks}\label{sec8}
Freedman's papers answer important questions about the properties
of OLS adjustment. He and others have summarized his results with
a ``glass is half empty'' view that highlights the
dangers of adjustment. To the extent that this view encourages
researchers to present unadjusted estimates first,
it is probably a good influence. The difference in means is the
``hands above the table'' estimate:
it is clearly not the product of a specification search,
and its transparency may encourage discussion of the strengths
and weaknesses of the data and research design.\footnote{On
transparency and critical discussion, see
\citet{AshPlant}, Freedman (\citeyear{Freed91}, \citeyear{Freed08c}, \citeyear{Freed10}),
Moher et al. (\citeyear{Moher}), and
Rosenbaum [(\citeyear{Rosenbaum10}), Chapter 6].}

But it would be unwise to conclude that
Freedman's critique should always override the
arguments for adjustment, or that
studies reporting only adjusted estimates
should
always be distrusted.
Freedman's own work shows that
with large enough samples and balanced two-group designs,
randomization justifies the traditional adjustment.
One does not need to believe in the classical linear model to tolerate or
even advocate OLS
adjustment, just as one does not need to believe in
the Four Noble Truths of Buddhism
to entertain the hypothesis that mindfulness meditation
has causal effects on mental health.

From an agnostic perspective, Freedman's theorems are a major
contribution.
Three-quarters of a century after Fisher
discovered the analysis of covariance, Freedman deepened our
understanding of its properties by deriving the regression-adjusted
estimator's asymptotic distribution without assuming a regression
model, a constant treatment effect, or an infinite superpopulation. His argument
is constructed with unsurpassed clarity and rigor.
It deserves to be studied
in detail
and considered carefully.

\section*{Acknowledgments}
I am grateful to Jas Sekhon, Dylan Small,
Erin Hartman, Danny Hidalgo,
and Terry Speed
for valuable advice;
to Peter Aronow, Gus Baker, Erik Beecroft, Dean Eckles, Jed Friedman,
Raphael Kang, Pat Kline,
Alex Mayer, Justin McCrary, David McKenzie, Tod Mijanovich, Luke
Miratrix, Deb Nolan,
Berk \"{O}zler, Susan Paddock,
Cyrus Samii,
an anonymous associate editor,
and reading groups at Berkeley and Penn
for helpful discussions;
and to David Freedman for his generous help
earlier in my education.
Any errors are my own.

\begin{supplement}[id=suppA]
\stitle{Proofs}
\slink[doi]{10.1214/12-AOAS583SUPP}  %[doi,text={...}] - jei reikia suskaldyti doi
\sdatatype{.pdf}
\sfilename{aoas583\_supp.pdf}
\sdescription{Proofs of theorems, corollaries, and selected remarks.}
\end{supplement}

% imsref loaded by akundreckaite, 2012-09-04 12:20:30
% imsref loaded by akundreckaite, 2012-09-04 12:44:51
% imsref loaded by akundreckaite, 2012-09-04 13:45:20
%

\printaddresses

\end{document}